\documentstyle[12pt]{article}

\begin{document}
\begin{titlepage}
\begin{flushright}
{\boldmath $Y\!ukawa$ $Institute$ $Kyoto$}\hfill
YITP-95-14\\
January 1996
\end{flushright}
\vskip 2em
\begin{center}
{\LARGE On the String Actions \\
      for the Generalized Two-dimensional \\
      Yang-Mills Theories \par} 
\vspace{25mm}
{\large \lineskip .5em
Yuji Sugawara\footnote{e-mail address: sugawara@yukawa.kyoto-u.ac.jp}
\par} \vspace{15mm}
{\large {\sl Yukawa Institute for Theoretical Physics\\
              Kyoto University, Kyoto 606, Japan}}\footnote
           {Present adress:  Department of Physics, Osaka University
          Machikaneyama 1-1, Toyonaka, Osaka 560, Japan}
\end{center} \par
\vspace{20mm}
\begin{abstract}
We study  the structures of  partition
functions of the large $N$ 
generalized two-dimensional Yang-Mills theories ($gYM_2$) 
by recasting the higher Casimirs.
We clarify the appropriate  interpretations of them and
try to extend the Cordes-Moore-Ramgoolam's
topological string model
 describing the  ordinary $YM_2$   \cite{CMR} to those describing 
$gYM_2$. We present the expressions of the appropriate operators 
to reproduce the higher Casimir terms in $gYM_2$.
The  concept of   ''deformed gravitational descendants'' 
will be introduced for  this purpose.
\end{abstract}

\end{titlepage}

\newcommand{\gh}{\mbox{gh}}
\newcommand{\met}{\mbox{Met}}
\newcommand{\map}{\mbox{Map}}
\newcommand{\tr}{\mbox{Tr}}
\newcommand{\Om}{\Omega}
\newcommand{\df}{\stackrel{\rm def}{=}}
\newcommand{\co}{{\scriptstyle \circ}}
\newcommand{\de}{\delta}
\newcommand{\si}{\sigma}
\newcommand{\ep}{\varepsilon}
\newcommand{\om}{\omega}
\newcommand{\al}{\alpha}
\newcommand{\la}{\lambda}
\newcommand{\La}{\Lambda}
\newcommand{\De}{\Delta}
\newcommand{\vphi}{\varphi}
\newcommand{\Exp}{\mbox{Exp}}
\newcommand{\rank}{\mbox{rank}}
\newcommand{\dimn}{\mbox{dim}}
\newcommand{\bm}[1]{\mbox{\boldmath ${#1}$}}
\newcommand{\ca}[1]{{\cal #1}}
\newcommand{\lb}{\lbrack}
\newcommand{\rb}{\rbrack}
\newcommand{\rn}[1]{\romannumeral #1}
\newcommand{\msc}[1]{\mbox{\scriptsize #1}}
\newcommand{\dsp}{\displaystyle}
\newcommand{\scs}[1]{{\scriptstyle #1}}
\newcommand{\deebar}{\bar{\partial}}
\newcommand{\br}{\mbox{R}}
\newcommand{\bc}{\mbox{C}}
\newcommand{\bz}{\mbox{Z}}
\newcommand{\bq}{\mbox{Q}}
\newcommand{\bn}{\mbox{N}}
\newcommand {\eqn}[1]{(\ref{#1})}
\newcommand{\binomial}[2]{\left(\ba{c}#1\\#2\ea\right)}

\newcommand{\be}{\begin{equation}}\newcommand{\ee}{\end{equation}}
\newcommand{\bea}{\begin{eqnarray}} \newcommand{\eea}{\end{eqnarray}}
\newcommand{\ba}[1]{\begin{array}{#1}} \newcommand{\ea}{\end{array}}


{\em 1.}
It is an old problem
to understand  relationships between  Yang-Mills theory and  string theory 
  \cite{old}, especially for the two-dimensional case $(YM_2)$ \cite{old2}.
In early years of 90's  one great progress 
was given by 
Gross and Taylor  within the framework of $YM_2$
on any compact Riemann surface $M$  \cite{GT}.
In these celebrated works
they investigated in detail the large $N$ expansion of 
the partition function by making use of 
some group theoretical techniques,
and showed that   it  is realized as an asymptotic 
series of which terms are all some homotopy invariants of 
ramified covering maps onto  the considered  
Riemann surface $M$. This strongly suggests the possibility of
reformulating  $YM_2$ as a theory of topological string with  
the space of maps $f ~:~\Sigma ~\longrightarrow ~M$ as the
configuration space.

Inspired  with these Gross-Taylor's studies, 
Cordes, Moore and Ramgoolam gave an elegant   
implication \cite{CMR}.
They firstly considered the 0-area limit (or equivalently,
 the weak coupling limit) of the large $N$ $YM_2$
and proved that each  term of the Gross-Taylor's 
asymptotic series is exactly equal to the Euler number of the moduli
space of branched covers.
Based on this observation they constructed the  world sheet action
of a topological string model corresponding to the large $N$ $YM_2$.
Their topological string theory is formulated to calculate
the Euler number of the moduli space and, to this aim, includes
some extra degrees of freedom - the ''co-fields'' \cite{CMR} (see also
\cite{vw}).
For the case of non-zero area,
they  gave  a conjecture (and partially proved)  
that by adding the simple   perturbation of the  
''area operator''  
one  can reproduce  the result of the non-zero area case.
They also presented a stimulating speculation;
the appropriate  
perturbations of the gravitational descendants of area operators
might correspond to the ''generalized 2-dim Yang-Mills theories''
($gYM_2$) \cite{W,GSY}, which are  given by replacing
the 2nd Casimir with some higher Casimirs 
in the heat kernel Boltsmann weight of $YM_2$ \cite{Mig,Rus,W}.
  
In this article 
we shall  present a detailed study of $gYM_2$ motivated 
with this speculation. 
We shall  exhibit
the world-sheet actions for the topological string models describing 
$gYM_2$, in other words, construct the suitable perturbation terms to the CMR's
string model reproducing the higher Casimirs. To this aim we shall establish 
 the rule to translate the algebraic data of higher
Casimirs appearing in $gYM_2$ into some geometric data fitted for  the topological 
string theory.   Our suitable perturbation terms will be expressed by
the {\em deformed\/} gravitational descendants, which will be defined later.

~

{\em 2.}
We shall start with a short review for the work \cite{CMR}.
Let $M$ be an arbitrary compact Riemann surface with genus $p$,
and $G$ be a compact group (gauge group).
The two-dimensional pure Yang-Mills theory
($YM_2$) on $M$ is usually defined by
\be
 S_{YM_2}(A_{\mu})=\frac{1}{4g^2}\int_M \, dv \, \tr (F^{\mu \nu}
F_{\mu \nu}) , \label{ym2action1}
\ee 
or by an equivalent form; 
\be
 S_{YM_2}(A_{\mu}, \phi)=\frac{1}{2}\int_M \, i \tr (\phi F) 
  + \frac{g^2}{4} \int_M \, dv \, \tr  (\phi^2). \label{ym2action2}
 \ee
Here $\phi$ is a scalar field valued in the Lie algebra of $G$.
Such a reformulation  of $YM_2$  
makes it  clearer that  $YM_2$ can be regarded as   
a topological gauge theory (the $BF$-theory) 
$\dsp S_{BF}= \frac{1}{2}\int_M \, i \tr (\phi F)$ with the perturbation 
term $\dsp \frac{g^2}{4} \int_M \, dv \, \tr (\phi^2) $
which breaks down the topological invariance.
Some detailed investigations from this point of view 
are given in \cite{W}.

The partition function of $YM_2$  is calculable  by the lattice 
method \cite{Mig,Rus} or by the continuum approach
\cite{W}. The exact result is  summarized as the  following
famous formula;
\begin{equation}
Z_{YM_2}(M, G)= \sum_{R} \,(\dim R)^{2-2p}\, e^{-\frac{g^2}{2} A C_2(R)}
\label{partition}
\end{equation}
where the summation is taken over all the equivalence classes of 
 irreducible representations of
$G$ and $C_2(R)$ is the 2nd Casimir operator.

Utilizing  some group theoretical techniques
Gross and Taylor derived  the formula of 
$1/N $ asymptotic expansion of this partition function
\eqn{partition}
for the case of $G = SU(N)$ with the redefinition of coupling 
constant $\la = g^2 N$ \cite{GT}. 
After that Cordes, Moore and Ramgoolam   made a refinement of  
the Gross-Taylor's formula \cite{CMR}. 
This can be written in the following way
only for the chiral sector\footnote{The meaning of ''chiral'' is given
in \cite{GT,CMR}.};
\be
\ba{l}
\dsp Z_{YM_2}^+(M, SU(N))  =1+  \sum_{n=1}^{\infty}\sum_{B=0}^{\infty}
   \,\left(\frac{1}{N}\right)^{n(2p-2)+B} \, e^{-\frac{1}{2}n \la A} \,
   e^{\frac{n^2}{2N^2}\la A} 
\,  \sum_{k=0}^{\infty} \, \frac{1}{k!}(-\la A)^k 
\, \sum_{L=0}^{\infty}\, \chi (\ca{C}_{L} (M))    \\
\dsp    \hspace{4cm} \times  
\sum_{v_1,  \ldots ,  v_{L} \in S_n \setminus \{1 \}}\, 
\sum_{p_1, \ldots , p_k \in T_2 }
\, \sum_{s_1,t_1 ,\ldots , s_p,t_p \in S_n } \\
\dsp \hspace{4.5cm} 
\times \, \frac{1}{n !} \delta (v_1 \cdots v_{L} \, p_1 \cdots  p_k
\, \prod_{i=1}^{p}s_it_is_i^{-1}t_i^{-1}) \, 
\delta_{\sum_{i=1}^{L} (n-K_{v_i}) \, +k ~, B} ~~~~  .
\ea \label{cgts2}
\ee
In this expression  we set 
\begin{equation}
\ca{C}_r (M) = \{ (z_1, ~\ldots, ~z_r ) \in M^r ~; ~ 
   z_i \neq z_j ~(\forall i\neq j )~\}/ S_r ~~~,
\end{equation}
where $S_r$ acts as the permutations of $z_1, ~ \ldots, ~ z_r$, and 
so we can explicitely write down its Euler number; 
\be
\dsp \chi (\ca{C}_r (M)) = 
\frac{(2-2p)(2-2p-1) \, \cdots \, (2-2p-r+1)}{r !} .
\ee
$K_v$ is the number of cycles in the cycle decompositon of   $v\in  S_n$,
$\delta (\cdots ) $ means the delta function on $\bc \lb  S_n \rb$ 
(the group algebra ove the symmetric group $S_n$) and 
$T_2 \subset S_n$
denotes the conjugacy class of the transpositions.

We survey  their  main results. 
For this purpose 
we shall neglect the factor $e^{\frac{n^2}{2N^2}\la A}$ for the time
being. 
Rewrite \eqn{cgts2} as 
$\dsp Z_{YM_2}^+(M, SU(N))  =  1 + \sum_{n=1}^{\infty}\sum_{B=0}^{\infty}
   \, \left(\frac{1}{N} \right)^{2h-2} \, e^{-\frac{1}{2}n \la A}\,
Z^+ (h,n, \la A)$ with the relation
$2h-2 = n(2p-2)+B$ (Riemann-Hurwitz).
We then find  that 
 $Z^+ (h, n, \la A)$ is a summation of some topological invariants of 
the $n$-fold branched covers $f \,  :\, \Sigma  \longrightarrow \, M$
$(h = \mbox{genus of }\Sigma)$, or equivalently, of the holomorphic mappings
with degree $n$. For the case of $ \la A=0$
they  proved  a remarkable fact;
\be
  Z^+ (h , n, 0) = \chi_{\msc{orb}} (\ca{F}_{h,n}) ,
\label{euler}
\ee
namely, $Z^+ (h , n, 0)$ is equal to the orbifold Euler number
of the moduli space of branched cover.
Based on this fact they constructed a model of topological string
such that it has  $Z^+(h,n,0)$ as its partition function.
They also conjectured  and partially proved
that the case of $\la A \neq 0$ is recovered by
including the perturbation of the ''area operator''
$\dsp \sim -\frac{\la}{2}\int_{\Sigma} \, f^* \om$, where
$\om $ is the volume form of $M$,
with carefully treating  the contact terms
in the  similar manner as those of topological gravity \cite{VV}.

Let us return to the formula \eqn{cgts2}.
 In the summation in the formula \eqn{cgts2}
the factors
$\dsp \frac{1}{k!}(-\la A)^k $
and $\chi (\ca{C}_{L} (M)) $  
correspond to the integrals over the continuous moduli.
On the other hand,
each of the elements of $S_n$ represents the combinatorial data of
the branched cover, namely, the summation over the elements of 
$S_n$ can be translated into the summation over the homotopy classes
of branched covers.
The delta function factor $\dsp \frac{1}{n!} \delta(v_1 \cdots)$
imposes the consistency condition to reconstruct the covering
Riemann surface.
The elements $v_1, \ldots, v_{L},$ $p_1, \ldots , p_k \, \in S_n$
respectively correspond to the branch  points, say,
$V_1, \ldots, V_{L},$ $P_1, \ldots , P_k$
 on $M$. Their correct contributions to the Euler number of the world
sheet $\Sigma$ are  equal to  
$\dsp n-K_{v_i}$  for $V_i$  $(i=1,\ldots,L)$ and $1$ for each $P_j$
$(j=1,\ldots, k)$. (This means that all $P_j$'s are simple branch points.)
The appearance of the factor
$\delta_{\sum_{i=1}^{L} (n-K_{v_i}) \, +k ~, B} $ ensures the correct
genus expansion of $\Sigma$.

We lastly comment on the role of the factor
$e^{\frac{n^2}{2N^2}\la A}$ which we neglected above.
A nice interpretation of this factor is presented  in \cite{GT},
namely, the ''infinitesimal tubes and handles''.
But  we have not succeeded in properly treating these objects and their 
generalizations for  $gYM_2$ in the framework of the CMR's topological string.
Fortunately, the contributions of these terms are absent for the case
of $G=U(N)$.
Hence we shall focus on  the $U(N)$-theory in the following discussions.

~

{\em 3.}
Now let us consider the ''generalized two-dimensional Yang-Mills 
theory''. This is defined by
\be
 S_{gYM_2}(A_{\mu}, \phi)= S_{BF}(A_{\mu},  \phi) +
  \int_M \, dv \, V  (\phi), \label{gym2action}
 \ee
where the ''potential'' $V(\phi)$ is an arbitrary  invariant polynomial. 
The gauge theory of this type was studied in \cite{W,GSY}.
The quantization of this theory is   completely parallel to
$YM_2$ and its  partition function is   given by the similar formula 
as \eqn{partition}. The only difference is that 
the  quadratic Casimir $C_2(R)$ appearing in this formula, 
which corresponds
to the term $\dsp \sim \int \, dv \, \tr (\phi^2)$ in \eqn{ym2action2}, 
is replaced with some higher Casimir operator corresponding to the general
potential $V(\phi)$ \cite{W,GSY}.

Consider the case; $\dsp V(\phi)= g_n \, \tr (\phi^n) +
\mbox{ lower terms}$. The corresponding Casimir operator generally
has the form;
$\dsp  g_n \, C_n(R) + \sum_{\{ k_i \}} \, a(k_1,k_2, \ldots) 
\prod_r \, (C_r(R))^{k_r}$,
where the summation of the 2nd term is taken over the set  
$\{ k_1, k_2, \ldots \}$ such that $\dsp \sum_r \, rk_r < n$
and $C_r(R)$ denotes the $r$-th Casimir operator of $U(N)$\footnote
    {Our definitions of the higher Casimirs  are slitely different 
   from those in \cite{GSY}. This difference reduces to the choice of
    the renormalization condition.};
\be
C_r(R)
 = \sum_{i=1}^N \, m_i^r \, \prod_{j\neq i}\, \left(
  1- \frac{1}{m_i -m_j}\right) .
\label{casimir}
\ee
In this expression we set $m_j =n_j + N-j$ with 
$(n_1, \ldots , n_N)$ being the signature of $R$. 
In order to determine the coeffecients $a(k_1,k_2, \ldots)$
from $V(\phi)$ we must fix the renormalization condition explicitely
(fix the defintion of the normal ordering). This is generally a
very complicated procedure and beyond the scope of this article. 
Instead   we shall simply {\em define\/} the $gYM_2$ 
by the following formula;
\begin{equation}
Z_{gYM_2}(M, G)= \sum_{R} \,(\dim R)^{2-2p}\, 
e^{-\sum_{r=1}^n \, g_r A C_r(R)\, +\, 
\sum_{\{ k_i \}} \, a(k_1,k_2, \ldots) 
\prod_s \, (C_s(R))^{k_s}}
\label{partition2}
\end{equation}
We especially focus on the cases such that only the linear terms 
$\dsp \sim \sum_{r=1}^n \, g_r A C_r(R)\,$ are included in 
the ''energy eigen-values'' of $gYM_2$ and will treat  the chiral
sector only.

Before developing the main discussions, we  
need to give a few remarks:
Firstly, the coupling constants $g_r$ should be rescaled 
as $\dsp g_r = \frac{\la_r}{N^{r-1}}$ when taking the large $N$-limit.
We will observe that 
these rescalings are necessary for the correct genus expansions.
We must also remark the following fact;
the $U(N)$-theory has 
an extra degrees of freedom, the ''$U(1)$-charge'', compared with
the $SU(N)$-theory. The irreps of $U(N)$ are classified 
by the $U(1)$-charge $Q$ and the irreps of $SU(N)$-subgroup
(Yang tableaus). Here we shall only consider the sector with
$Q=0$ in order to make things easy.
This has the similar structure 
as the $SU(N)$-theory, but  easier to treat, 
since the higher Casimirs of $U(N)$ is much  simpler than 
those of $SU(N)$. As has been already pointed
out, the $SU(N)$-theory
includes some extra terms 
which are the natural generalizations of  
the infinitesimal tubes and handles
in the case of $YM_2$. 
We would like  to argue on the sectors of non-zero $Q$ and
on the $SU(N)$-theory elsewhere.

Now,
the question we want to solve is as follows:
What is the topological string action corresponding to the model
\eqn{partition2}? 
In other words, what is the perturbation terms to the CMR's string
action which recover the contributions of the higher casimirs 
$\dsp \sim \sum_{r=1}^n \, g_r A C_r(R)$? 
To search for the solution of this problem let us proceed along the
same line from \eqn{partition} to \eqn{cgts2}.
So, we need to clarify  the $N$-dependence of $C_r(R)$.
Set $R \in Y_n$ 
(the set of Young tableaus with $n$-boxes)
and assume that $n << N$, since we are considering the chiral sector only.
It is convenient to  expand  formally $C_r (R)$ with respect to $N$;
\be
  \frac{1}{N^{r-1}}C_r (R) = \sum_{l=1}^r \, \binomial{r-1}{l-1}
        \frac{1}{N^{l-1}}\Xi_{l}(R) .
\label{Xi}
\ee
Here the coefficients $\Xi_{l}(R)$ are defined not to depend on $N$. 
One remarkable fact is that 
$\Xi_{l}(R)$'s  so defined are  actually  independent of
the value of $r$. The binomial coefficients
 $\binomial{r-1}{l-1}$ are suitably 
chosen to assure of this property.
In the similar manner as \cite{GT,GSY} we can  represent  $C_r (R) $,
or rather $\Xi_l (R)$, in terms of the language of symmetric group.
After some calculations we can obtain
\be
\ba{lll}
\Xi_1(R) & = & n \\
\dsp \Xi_2(R) & = & \dsp 2 \frac{\chi_R (P_2)}{d_R} \\
\dsp \Xi_3(R) & = & \dsp 3 \frac{\chi_R (P_3)}{d_R} + n(n-1) \\
\dsp \Xi_4(R) & = & \dsp 4 \frac{\chi_R (P_4)}{d_R} + (6n-10) 
   \frac{\chi_R(P_2)}{d_R} \\
\dsp \Xi_5(R) & = & \dsp 5\frac{\chi_R (P_5)}{d_R} + 
    3(4n-7)\frac{\chi_R(P_3)}{d_R} \\
  & &\dsp  ~~~+16\frac{\chi_R(P_{2,2})}{d_R} + 2n(n-1)(n-2) + n(n-1) \\
 \cdots &&
\ea
\label{Xi example}
\ee
In these expressions 
$\chi_R(\cdots)$, $d_R$ mean the character and the dimension 
as the irrep of $S_n$.\footnote
   {Accoding to the convention of \cite{GT}, we often use the letter $R$
    either  as  the meaning of the irrep of $U(N)$ ($SU(N)$) or
     the meaning of  
     irrep of $S_n$ when they are expressed by the same Young
    tableau. It may be somewhat confusing,
     but we dare to do so to avoid the notational complexities.}
 $P_{n_1,n_2,\ldots} (\in \bc \lb S_n \rb )$ denotes the sum
of all the elements of $S_n$ belonging to the conjugacy class 
represented by $\{n_1 , n_2 , \ldots \} \in Y_n$, and we employ 
some abbreviations, say, $P_2$ truly means 
$P_{\scs 2,\underbrace{1,\ldots,1}_{\msc{$n-2$ times}}} \equiv
$ sum of all transpositions.
It may be  convenient to define the elements  
$\Xi^{(n)}_l \in\mbox{Center}(\bc \lb S_n \rb)$  
such that $\Xi_l(R)=\rho_R(\Xi^{(n)}_l) 
\equiv \dsp \frac{\chi_R(\Xi^{(n)}_l)}{d_R}$
where $\rho_R(\cdots)$ means the representation matrix.
For example, $\Xi^{(n)}_2 = 2P_2$, $\Xi^{(n)}_3 =3 P_3 + n(n-1)$,
 $\ldots $ and so on.

In this way we have succeeded in translating  the language   
of $U(N)$ into that of $S_n$.    It is easy to write down  
the similar formula as \eqn{cgts2}  for   $gYM_2$  from the formulas
\eqn{Xi example}.
We should recall that 
this procedure  is a cornerstone
of the stringy interpretations of $YM_2$ \cite{GT,CMR}.
So, we can already give the similar interpretations of $gYM_2$
in the {\em qualitative\/} level. In fact these subjects are 
carefully discussed in \cite{GSY}.
To solve our question,  however, we need more {\em quantitative\/}
informations. We  shall  restart with  the following identity on 
$\bc \lb S_n \rb$;
\be
\Xi^{(n)}_l = \sum_{a=1}^n \, \sum_{b_1, \ldots ,b_{l-1} \neq a} \,
 p_{ab_1} \cdots p_{ab_{l-1}} , 
\label{Xi formula}
\ee
where $p_{ab}( \equiv (ab)~)$ denotes
the transposition acting on the $a$-th and $b$-th elements.
This can be  proved by simple observations about the definition of 
$\Xi^{(n)}_l$.
This identity \eqn{Xi formula}  includes the sufficient informations we want and leads to 
a manifest  interpretation; 
{\em $\Xi^{(n)}_l$ represents 
the degeneration of $l-1$
simple ramification points.}  
Here we should notice that  the factors $(1/N)^{l-1}$ of  
$\Xi^{(n)}_l$  appearing in  the identity  \eqn{Xi}  completely match with this interpretation.
In fact the existence of $l-1$ simple ramification points 
adds $-(l-1)$ to the Euler number
of the world sheet and therefore  these factors 
give the correct genus expansion.
If we adopt the rescaling  $\dsp g_r =\frac{\la_r}{N^{r-1}}$, which we mentioned before,
all the terms $\Xi_l^{(n)}$ will appear in our equations with the suitable 
factors $(1/N)^{l-1}$.
It  is an interesting feature  that the $N$-dependences of all the coupling constants
are decided from the appropriate stringy interpretaion.    It might be analogous to
the situation of the doubly scaled matrix models \cite{matrix}.

To treat our problem more concretely we must prepare some mathematical
objects.
Let $\ca{F}_{h,n,s}$ be the  moduli space
of the $n$-fold branched cover over $M$ with genus $h$ 
and $s$-punctures, in other words, the moduli space of toplogical
string whose target space is $M$ ($n$ is usually called the 
''instanton number'' or the ''winding number'' 
in  the language of string theory);
\be
\ba{lll}
\ca{F}_{h,n,s} &\df& \{ (\, \Sigma, ~f, ~ x_1,~\ldots , ~x_s \,) ~:~
    f \, : \, \Sigma \, \longrightarrow \, M ~ \mbox{$n$-fold branched
cover, }\\
 &&  \hspace{1in} ~~~  x_1, \ldots, x_s \in \Sigma ~
\}/ \mbox{Diff} ~~,
\ea
\ee
More precisely we must designate  a suitable compactification
rule. We shall here adopt the compactification appearing in \cite{CMR}
with respect to the collisions of ramification points,
and the stable compactification with respect to the collisions of
punctures.
We also use the abbreviated notation $\ca{F}_{h,n}$,
which we  used in the previous discussions, 
instead of $\ca{F}_{h,n,0}$. 

Consider the one-puncture case $\ca{F}_{h,n,1}$.
Roughly speaking, $\ca{F}_{h,n,1}\sim \ca{F}_{h,n}\times \Sigma$,
so we can obtain the fibration;
\be
\pi ~:~ \ca{F}_{h,n,1} ~\longrightarrow ~ \ca{F}_{h,n},
\label{pi}
\ee
such that the fiber  $\pi^{-1}(\lb \Sigma, f  \rb) \cong \Sigma$ 
describes  the position of puncture.

For a fixed branched cover $f \, :\,  \Sigma \, \rightarrow \, M$,
we define the ''ramification divisor'' of $\Sigma$ by
\be
R_{\Sigma, f} \df \sum_{p \in \Sigma} \, (e(p)-1)p , 
\label{R}
\ee
where $e(p)$ is the ramification index of $p\in \Sigma$.
(The summation is well-defind since $e(p)=1$ for the unramified points.)
The divisor $R_{\Sigma,f} \subset \Sigma$ naturally induces 
a divisor $\ca{R}$ of $\ca{F}_{h,n,1}$ defined by;
\be
\ca{R} \cap \pi^{-1}(\lb \Sigma, f \rb)  = \lb R_{\Sigma , f}\rb ,
\label{caR}
\ee
where $\lb \cdots \rb$ means the isomorphism class of divisor 
defined by the diffeomorphisms.
Heuristically, 
$\ca{R}$ is nothing but the subvariety of $\ca{F}_{h,n,1}$
composed of the configurations such that the puncture collides with
some ramification points. 
Let us  denote the holomorphic line bundle 
corresponding to $\ca{R}$ by $\ca{L} ~\rightarrow ~ \ca{F}_{h,n,1}$.
More explicitely $\ca{L}$ is the line bundle whose fiber is given by
\be
   \ca{L}_{\lb \Sigma , f, x  \rb}  = K_{\Sigma} \otimes f^*(TM)|_x ,
\label{fiber L}
\ee
where $K_{\Sigma}$ means the canonical bundle of $\Sigma$,
$TM$ means the holomorphic tangent bundle of $M$.
The simplest  (somewhat heuristic) explanation that 
the line bundle $\ca{L}$ having this fiber indeed corresponds to 
the divisor $\ca{R}$ is as follows:
Consider the holomorphic  section
\be
 s ~: ~ \lb \Sigma, f, x \rb \in \ca{F}_{h,n,1}
 ~\longmapsto ~ \frac{df}{dz}(x) \in K_{\Sigma} \otimes f^*(TM)|_x . 
\ee
Clearly the zero-locus of this section coincides with the subset
of configurations such that the puncture $x$ coincides with a
ramification point, that is, $\ca{R}$ itself.

Let us return to the main discussions.
Our goal is to construct the suitable observables  
reproducing the higher Casimir terms $\sim \sum_{r=1}^n \, g_r A C_r(R)$
in \eqn{partition2} in the framework of the CMR's string theory.
We first introduce the ''area-operator'';
\be
\ca{A} \df \int \, f^* \om ,
\label{area}
\ee
where $\om$ means the volume form on $M$. 
As is well-known \cite{VV,witteng}, 
we can equivalently  use the 0-form component folded with 
the puncture operator $P$;
\be
\ca{A} = \om_{ij}(f(x))\, \chi^i\chi^j \, P .
\label{area2}
\ee
where $\chi^i$ denotes the usual ghost field of the topological 
$\sigma$-model.
Consider the operator of the form 
$(c_1(\ca{L}))^l \, \ca{A}$.
It is analogous to the gravitational descendants of the theory
of topological gravity \cite{witteng,VV}. In that case
one considered  the line bundle whose fiber is
\be
\ca{L}^{\msc{top grav}}|_{\lb\Sigma , f, x\rb} = K_{\Sigma}|_x ,
\ee
instead of \eqn{fiber L}.
Therefore it may be  reasonable  to  define these objects 
as the {\em deformed\/} gravitational descendants of the area 
operator.
Let us  express the form of the operator 
$c_1(\ca{L})$  more explicitely.
Recalling the structure of fiber of $\ca{L}$
\eqn{fiber L}, we can immediately find that 
\be
c_1(\ca{L}) = c_1(\ca{L}^{\msc{top grav}}) + R_{ij}(f(x))\, \chi^i
\chi^j .
\label{explicit}
\ee
The first term in the R.H.S
corresponds to the usual gravitational descendant 
(the operator often written as 
''$\gamma^0$'' in the several papers
of topological gravity), and the second  term  means nothing but 
the curvature two form on the target space $M$. 

In order to clarify the geometrical meanings of these operators 
let us consider the Poincare dual of them.
First we notice that 
\be
\ba{lll}
\mbox{Poincare dual of }\ca{A} &\cong& \ca{P}_y \\
  & \equiv &  \{ \mbox{configurations such that the puncture }
       \\
  &&\hspace{1in} \mbox{ is mapped to $y \in M$ } \} ~~~,
\ea
\label{pdA}
\ee
with a some fixed point $y \in M$.
It is also convenient to introduce
\be
\ba{lll}
\ca{R} & = & \ca{R}_1 \supset \ca{R}_2 \supset \ca{R}_3 \supset
  \cdots   \\
\ca{R}_l & = & \{ \mbox{configurations such that
 the puncture collides with the  point} \\
  & & \hspace{1in} \mbox{  which is  
  a degeneration of $l$ simple ramification points} ~\} .
\ea
\label{Rl}
\ee
Clearly $\mbox{codim} \, \ca{R}_l = l$ in $\ca{F}_{h,n,1}$.
Recalling that $c_1(\ca{L})$ is the Poincare dual of the divisor
$\ca{R}$, we now obtain that 
\be
\mbox{Poincare dual of } (c_1(\ca{L}))^l \, \ca{A}
  \cong  \ca{R}_l \cap \ca{P}_y .
\label{Rl2}
\ee
This represents the geometrical meaning of 
$(c_1(\ca{L}))^l \, \ca{A}$ and suggests the relation with 
$\Xi^{(n)}_{l+1}$ introduced above.

There is still one thing that we should remark:
Since the CMR's topological string theory has the vanishing
ghost number anomaly, it is covenient to treat only the operators 
with the vanishing ghost numbers.
Therefore, we shall rather consider  the operator 
$\ca{C}^l \, \ca{A} $ with the definition
\be
 \ca{C} \df c_1 (\ca{L}) \, (\hat{\bm{A}}, \Pi_0 \, \hat{\bm{A}}) ,
\label{C}
\ee
instead of $(c_1(\ca{L}))^l \, \ca{A}$ itself.
Here $\hat{\bm{A}}$ denotes the ''co-anti ghosts'' \cite{CMR},
which is, in some sense, the dual of the ghost fields and has the same
spin contents as those of the ghosts.
$\Pi_0$ is the projector onto the space of zero-modes.
This is actually a BRST-invariant object. The projector 
$\Pi_0$ is needed to assure this invariance for the factor
$(\hat{\bm{A}}, \Pi_0 \, \hat{\bm{A}})$,  of which role is merely to cancel 
the ghost number of $c_1 (\ca{L})$.

We are now in a position to demonstrate the main results of this article.
Let us write the correlator of the CMR's topological string 
defined on the world sheet  $\Sigma$ with genus $h$ as
\be
\dsp \langle \cdots \rangle_h  \equiv 
\int \, D(\bm{F},\hat{\bm{F}},\cdots ~) \, \cdots \,
e^{-I_{\msc{CMR}}(\bm{F},\hat{\bm{F}},\cdots ~)} ,
\label{correlation}
\ee
and similarly express the contribution from the $n$-instanton sector
by $\dsp \langle \cdots \rangle_{h,n}$.
One of the main results of \cite{CMR} can be written as   
\be
\langle\, 1\, \rangle_{h,n}
= \chi_{\msc{orb}}(\ca{F}_{h,n}).
\ee
First we should notice the following relation;
\be
\langle \ca{A} \rangle_h = \sum_{n=1}^{\infty}\,  
nA \langle 1 \rangle_{h,n}  + \langle \ca{C} \, \ca{A} \rangle_h .
\label{A1}
\ee
The second term in the R.H.S represents  the contact terms between
the area operator $\ca{A}$ and the ramification points   
which  are discussed in \cite{CMR}.
We may also rewrite it as follows;
\be
\ba{lll}
\langle (1-\ca{C})\, \ca{A} \rangle_h &=& \dsp \sum_{n=1}^{\infty} \, 
nA \langle 1 \rangle_{h,n}    \\
  & = & \dsp
  \ A \, \sum_{n=1}^{\infty} \, 
\sum_{L=0}^{\infty}\, \chi (\ca{C}_{L} (M))   
\sum_{v_1, ~ \ldots , ~ v_{L} \in S_n \setminus \{1 \}}\, 
\, \sum_{s_1,~t_1 ,~\ldots ,~ s_p,~t_p \in S_n }  \\
&& \dsp ~~~~~~\times
\, \frac{1}{n !} \delta ( \Xi^{(n)}_{1} \, v_1 \cdots v_{L} 
\, \prod_{i=1}^{p}s_it_is_i^{-1}t_i^{-1}) \, 
\delta_{\sum_{i=1}^{L} (n-K_{v_i}) \, , B(h,n)}   ~,
\ea
\label{A2}
\ee
where $B(h,n)$ is defined by $2h-2 = n(2p-2) +B(h,n)$.
Remembering the formulas \eqn{Xi formula}, \eqn{Rl2}
and the definition of $\ca{R}_l$, we can further obtain the analogous
identity for the deformed gravitational descendants; 
\be
\ba{lll}
\langle \, (1-\ca{C})\, \ca{C}^l \, \ca{A}\,  \rangle_h &=&  \dsp
  \ A \, \sum_{n=1}^{\infty} \, 
\sum_{L=0}^{\infty}\, \chi (\ca{C}_{L} (M))   
\sum_{v_1, ~ \ldots , ~ v_{L} \in S_n \setminus \{1 \}}\, 
\, \sum_{s_1,~t_1 ,~\ldots ,~ s_p,~t_p \in S_n }  \\
&& \dsp ~~~~~~\times
\, \frac{1}{n !} \delta ( \Xi^{(n)}_{l+1} \, v_1 \cdots v_{L} 
\, \prod_{i=1}^{p}s_it_is_i^{-1}t_i^{-1}) \, 
\delta_{\sum_{i=1}^{L} (n-K_{v_i}) + l \, , B(h,n)}   .
\ea
\label{A3}
\ee
The factor $(1-\ca{C})$  
suppresses again the extra contributions from the contact terms. 
If we incorpolate the suitable factors $(1/N)^{l}$ for  $\Xi^{(n)}_{l+1} $,
we can find out the  simple correspondence;
\be
  (1-\ca{C})\, \ca{C}^l  \, \ca{A} ~
\longleftrightarrow ~\, A \,  \left(\frac{1}{N}\right)^l   \, \Xi^{(n)}_{l+1} .
\label{A4}
\ee
Recalling  the relation of the Casimir operators and $\Xi_l^{(n)}$
\eqn{Xi},  we  finally obtain 
\be
\ba{l}
\dsp (1-\ca{C})(1+\ca{C})^l\, \ca{A} \equiv 
  \sum_{r=0}^{l}\,  \binomial{l}{r} \, 
(1-\ca{C})\, \ca{C}^r \, \ca{A} ~ \\
\dsp \hspace{1.5cm}~~
\longleftrightarrow  ~
 A \, \sum_{r=0}^l \, \binomial{l}{r} \, \left(\frac{1}{N}\right)^r \,   \Xi^{(n)}_{r+1} ~
\longleftrightarrow ~ A\,  \left(\frac{1}{N}\right)^l    \,   \,C_{l+1}(R) .
\ea
\label{A5}
\ee
 In this way  we have arrived at the solution of our main question:
{\em $gYM_2$ including the higher Casimir 
terms $\dsp  \sim A \,\sum_{r=1}^m \, \frac{\la_r}{N^{r-1}} \, C_r (R)$
can be recovered by the CMR's 
topological string with the perturbation term\/}
\be
\dsp \sum_{r=1}^m \, \la_r \, \int \, 
(1-\ca{C}) (1+\ca{C})^{r-1} \, \ca{A}. 
\ee

To close our discussions let us present a few comments:
Firstly,  notice  that 
the area-area contact terms vanish \cite{CMR}.
Hence the results  \eqn{A1} \eqn{A2}  \eqn{A3} can be 
exponentiated with no difficulty  and one can obtain the  correct   
polynomials of $A$.    This  immediately leads to
our final results.

Secondly, let us consider 
the special case for $C_2(R)$ -  the case of
usual $YM_2$. According to the
above rule  we obtain as the appropriate perturbation term
$  \dsp \sim \, \la_2 \, \int \, (1-\ca{C}^2) \, \ca{A} $.
This does not coincide  with the CMR's original conjecture;
$\dsp \sim \, \la_2 \, \int \, \ca{A}$.
But their arguments are those  about only the simple Hurwitz space 
($=$ the subspace of $\ca{F}_{h,n}$ composed of the configurations
such that all the ramification points are simple).
Hence our result is compatible with theirs, since the correction term
$-\ca{C}^2 \, \ca{A}$ does not contribute to the simple Hurwitz
space. 

~

{\em 4.} 
In this article we investigated the two-dimensional generalized 
Yang-Mills theories   from the point of view of the topological
string theory.
We have uncovered the geometrical meaning of the higher Casimirs
by recasting them, namely, by introducing the  elements 
$\Xi_l^{(n)}$ of $\bc \lb S_n \rb$.
Our most remarkable success 
is to present  the expressions of the perturbation
terms appropriate to describe the models of  $gYM_2$
including rather general higher Casimirs.
But it is {\em not\/} completely general.
We have not yet succeeded in treating
more general forms of higher Casimirs including the non-linear terms. 
We also have not succeeded in  treating  the sector with non-zero $U(1)$-charges,
and working  with the $SU(N)$-theory.
They are the open problems to be resolved, and I think, 
these problems might deeply relate with one another.
In order to overcome these problems
it will be necessary to 
develop our theory so that we can work with the terms of
''infinitesimal tubes and hundles'', which we neglected.

Another open problem  is  of course to  extend our results 
to the non-chiral case.  To this aim it may be more helpful to work 
in the framework of Ho\v{r}ava's theory (''topological rigid string'')
\cite{horava} than that of CMR's theory. His string model 
possesses  the moduli space of 
the minimal area maps rather than the (anti-)holomorphic
maps. Hence the non-chiral sector is naturally incorporated from 
the begining.

The problem that I think also interesting is as follows:
The CMR's studies disclosed the deep relation between
$YM_2$ (and $gYM_2$) and the Euler characteristic of moduli
space. On the other hand, the matrix models of 
the Kontsevich-Penner type \cite{KP} give the analogous results.
It is well-known that the matrix models of this type 
naturally give the simplicial decompositions of the moduli space
by the Feynmann diagrams.
It may be  meaningful  if we can  develop   the similar  diagramatic arguments 
for  $YM_2$ (and $gYM_2$).

\end{document}